\documentclass[11pt,a4paper]{article}
\usepackage{jcappub}
\usepackage{comment}
\usepackage{subfigure}
\usepackage{float}
\newcommand{\be}{\begin{equation}}
\newcommand{\ee}{\end{equation}}
\newcommand{\bea}{\begin{eqnarray}}
\newcommand{\eea}{\end{eqnarray}}
\newcommand{\Beq}{\begin{equation}\begin{aligned}}
\newcommand{\Eeq}{\end{aligned}\end{equation}}
\newcommand{\mpl}{M_{\rm Pl}}

\newcommand{\pr}{{\rm pr}}

\usepackage{color}
\usepackage{ifthen}
\newboolean{editorial}
\setboolean{editorial}{true}
\newcommand{\editorial}[2]{\ifthenelse{\boolean{editorial}}{\textcolor{red}{[\textsf{\textbf{{#1}}}: }\textcolor{blue}{\textsf{{#2}}}\textcolor{red}{]}}{}}

\usepackage{xcolor}

\begin{document}

\title{Scalar Radiation with a Quartic Galileon}

\author[a,b,c]{Claudia de Rham}
\author[d,b,e]{John T. Giblin, Jr.}
\author[a,b,c]{Andrew J. Tolley}

\affiliation[a]{Department of Physics, Blackett Laboratory, Imperial College, London, SW7 2AZ, UK}
\affiliation[b]{CERCA/ISO and Department of Physics, Case Western Reserve University, Cleveland, Ohio 44106, U.S.A.}
\affiliation[c]{Perimeter Institute for Theoretical Physics,
31 Caroline St N, Waterloo, Ontario, N2L 6B9, Canada}

\affiliation[d]{Department of Physics, Kenyon College, Gambier, Ohio 43022, U.S.A.}

\affiliation[e]{Center for Cosmology and AstroParticle Physics (CCAPP) and Department of Physics, The Ohio State University, Columbus, OH 43210, USA}

\emailAdd{c.de-rham@imperial.ac.uk}
\emailAdd{giblinj@kenyon.edu}
\emailAdd{a.tolley@imperial.ac.uk}

%\date{September 2023}
%\begin{abstract}

\abstract{The class of Galileon scalar fields theories encapsulate the Vainshtein screening mechanism which is characteristic of a large range of infrared modified theories of gravity. Such theories can lead to testable departures from General Relativity through fifth forces and new scalar modes of gravitational radiation. However, the inherent non-linearity of the Vainshtein mechanism has limited analytic attempts to describe Galileon theories with both cubic and quartic interactions. To improve on this, we perform direct numerical simulations of the quartic Galileon model for a rotating binary source and infer the power spectrum of given multipoles. To tame numerical instabilities we utilize a low-pass filter, extending previous work on the cubic Galileon. 
Our findings show that the multipole expansion is well-defined and under control.
Moreover, our results confirm that despite being a non-linear scalar, the dominant Galileon radiation is quadrupole, and we find a new scaling behaviour deep inside the Vainshtein region.
}
%\end{abstract}

\maketitle

\section{Introduction}
Of all the screening mechanisms proposed to account for how light dark energy degrees of freedom can affect cosmological evolution without being ruled out by current tests of gravity, the Vainshtein mechanism \cite{Vainshtein:1972sx,Deffayet:2001uk,Dvali:2002vf,Lue:2003ky,Babichev:2009us,Babichev:2009jt,Babichev:2010jd,deRham:2010ik,deRham:2010kj,Babichev:2013usa} remains simultaneously the most interesting and potentially technically natural but least well understood. It is automatically built into large classes of massive theories of gravity (both soft and hard mass) without any need for tuning \cite{Luty:2003vm,Babichev:2009us,Babichev:2009jt,Babichev:2010jd,deRham:2010ik,deRham:2010kj,deRham:2007xp,deRham:2007rw,Babichev:2013usa,Ondo:2013wka,Joyce:2014kja,deRham:2014wfa,deRham:2023byw}. Furthermore these theories have a universal decoupling limit whose description in terms of a scalar field accompanied by an intriguing non-linearly realised {\it{Galileon}} symmetry has an interesting mathematical structure \cite{Nicolis:2008in,deRham:2010eu,Goon:2011qf}. However, the Vainshtein mechanism is poorly understood beyond idealized symmetric situations due to its inherent non-linear nature, although various results have been obtained in a variety of contexts \cite{%Deffayet:2001uk,Babichev:2009us,Babichev:2010jd,
Babichev:2011iz,DeFelice:2011th,Kaloper:2011qc,Kimura:2011dc,Chkareuli:2011te,Gannouji:2011qz,Chu:2012kz,Sbisa:2012zk,Padilla:2012ry,Hui:2012jb,Hiramatsu:2012xj,Belikov:2012xp,Li:2013nua,Winther:2015pta,Berezhiani:2013dca,Barreira:2013eea,Ogawa:2018srw,Iorio:2012pv}. Furthermore, those non-linear interactions significantly complicate numerical evolution with certain approaches being ill-posed from the outset. 
\\

The simplest example of the Vainshtein mechanism is provided by the cubic Galileon \cite{Luty:2003vm} which originally emerged as the decoupling limit of the Dvali-Gabadadze-Porrati model \cite{Dvali:2000rv}. This model is sufficiently simple to be reasonably acquiescent to analytic approximations \cite{deRham:2012fw,deRham:2012fg}. In a recent work, three distinct numerical approaches were utilized to describe the scalar radiation emitted by a binary system, such as a binary pulsar \cite{Gerhardinger:2022bcw}. One of these approaches built on previous works \cite{Dar:2018dra} by directly simulating the cubic Galileon using a low-pass filter to suppress any numerical instabilities. The other approaches utilized an extended system of equations of motion, in effect a well-posed UV completion, designed to reproduce the Galileon theory at low energies. In the present work we will extend the low-pass filter approach to the quartic Galileon, and in a companion work \cite{Gerhardinger:2024} the UV completion method will similarly be utilized for the quartic Galileon theories. What makes the quartic Galileon particularly challenging is that unlike in the cubic case, analytic attempts to describe scalar radiation using the one-body effective approach fail \cite{deRham:2012fg}. This failure is a reflection of the fact that on linearizing around spherically symmetric backgrounds, the usual centrifugal repulsion which suppresses power in high order multipoles is lost, leading to a superficially divergent power spectrum. This simply indicates the failure of perturbative analytic approaches. It is worth noting that this problem is ameliorated by having a larger cubic Galileon term, and this is mirrored by our numerical simulations which are more stable for large cubic interactions.\\

In the present work we show that just as for the cubic case, the Vainshtein mechanism is successfully realized for these time-dependent solution in the quartic Galileon models (in the presence of a cubic interaction), something which could not be inferred from any approximate analytic treatment. In common with the pure cubic Galileon, we find that the dominant scalar radiation is quadrupole. However, when the source radiates in the region in which the quartic interaction dominates, the scaling of the radiated power with both frequency and multipole number is distinct from the cubic case demonstrating that the quartic Galileon screening while qualitatively similar to the cubic is quantitatively distinct. The results in this paper closely parallel those in \cite{Gerhardinger:2024} where the quartic Galileon system is treated by embedding it in a `UV completion' that renders the dynamics well-posed.  \\

\section{The Model}\label{sec:model}

The Galileon is a scalar field theory which exhibits a non-linearly realized `Galileon' symmetry $\pi \rightarrow \pi + v_{\mu} x^{\mu}$ \cite{Nicolis:2008in}. Among the class of interactions that are invariant under this symmetry are a finite set of interactions that lead to second order equations of motion. In $3+1$ dimensions these are the cubic, quartic and quintic Galileon terms. In the present work we shall be concerned with the generic quartic Galileon meaning that we shall include both cubic and quartic interactions and consider a coupling to the trace of the stress energy of matter\footnote{We closely follow the notation of \cite{Gerhardinger:2024}. }
\begin{equation}
	\label{eqn:eom}
		 \Box\pi + \frac{1}{3\Lambda_3^3}\left((\Box\pi)^2 - (\partial_\mu \partial_\nu \pi)^2 \right) + \frac{1}{9\Lambda_4^6} \left((\Box\pi)^3 - 3\Box\pi (\partial_\mu \partial_\nu \pi)^2 + 2 (\partial_\mu \partial_\nu \pi)^3 \right) = \frac{T}{3\mpl} \,.
\end{equation}
The precise form of the normalization is characteristic of how the Galileon arises as the helicity-zero mode in massive gravity theories \cite{deRham:2014zqa,deRham:2023byw}.
If we assume spherical symmetry and a static point-source, we can integrate \eqref{eqn:eom} to obtain an ordinary differential equation for $E \equiv d\pi/dr$ \cite{Nicolis:2004qq,deRham:2012fg,Berezhiani:2013dca}
\begin{equation}
	\label{eqn:ssymetry}
		\frac{E}{r} + \frac{2}{3\Lambda_3^3}\left(\frac{E}{r}\right)^2 +\frac{2}{9\Lambda_4^6}\left(\frac{E}{r}\right)^3= \frac{1}{12\pi r^3}\frac{M_s}{\mpl} \, .
\end{equation}
We will use a system consisting of two gaussian `masses' orbiting around their center of mass
\begin{equation}
        T = - A \left(e^{-\left({\vec{r}_+}(t)/\sigma\right)^2}+e^{-\left({\vec{r}_-}(t)/\sigma\right)^2}\right) \, .
    \end{equation}
The locations of each mass, at a given time, are $ {\vec{r}} (t) = \left(x \pm \bar{r}\cos\left(\Omega_p t\right)/2, y \pm \bar{r}\sin\left(\Omega_p t\right)/2, z\right)$ and the constant $A$ is chosen so that the total mass of the system is $M_{\rm s} = \int d^3x\, \rho = \int d^3 x \,T$ and $\bar{r}$ is the diameter of the circular orbit. We constrain the system to be Keplerian,
\begin{equation}
    M_s = 8\pi \mpl^2 \bar{r}^3 \Omega_p^2 \, ,
\end{equation}
for an orbital frequency $\Omega_p$, which is to say we will ignore relativistic corrections to the orbit.  The quantity $\Omega_p\bar{r}$ is an important quantity that parameterizes the power in scalar radiation from the Klein-Gordon system, 
\begin{equation}
    P_2^{\rm KG} = \frac{M_s}{\bar{r}}\frac{\left(\Omega_p \bar{r}\right)^8}{45}.
    \label{eq:kgpower}
\end{equation}

\section{Numerical Implementation}\label{sec:numerics}
The Galileon system has been studied both analytically and numerically (see e.g. \cite{deRham:2012fg,deRham:2012fw,Chu:2012kz,deRham:2012az,Dar:2018dra,Kaloper:2011qc,Iorio:2012pv,Barreira:2013eea,Ogawa:2018srw}). At least two of the numerical implementations use modified versions of {\sc GABE} using a standard, CPU implementation \cite{Dar:2018dra} as well as a GPU implementation \cite{Gerhardinger:2022bcw}.  In \cite{Dar:2018dra} the authors show that turning on the sources and the non-linear interactions slowly at the beginning of the simulation can avoid the issue of a singluar effective metric \cite{deRham:2012fg}.  This method was made more robust in the GPU implementation, one of the methods described in \cite{Gerhardinger:2022bcw}, where efficient Fast Fourier Transforms (FFTs) made it possible to calculate mixed spatial derivatives using spectral methods and cut-off high frequency modes. In the present work our goal is to extend these methods to the quartic Galileon. \\

All of the simulations presented here use grid sizes of $N^3=384^3$ points where the length of each side is $L = 50 \bar{r}$.  These parameters are chosen as to be comparable to the previous literature.
We use a set of dimensionless units to rescale spacetime (we can choose physical units such that $\mpl=1$)
\begin{equation}
    x^\mu_\pr = \frac{2}{\bar{r}} x^\mu  \, ,
\end{equation}
and the Galileon field,
\begin{equation}
    \pi_\pr = \sqrt{\frac{\bar{r}}{M_s}}\pi.
\end{equation}
In these dimensionless units, we can re-write the quartic equation of motion, \eqref{eqn:eom}, in a pictorial form that allows us to separate the non-linear interactions $\mathcal{O}_n$ from the strength of these interactions $\kappa_n$ and the numerical turn-on functions, $f_n$,
\begin{equation*}
	\label{eqn:dimlesseom}
		 \mathcal{O}_2+f_3 (t_\pr)\kappa_3 \mathcal{O}_3 + f_4 (t_\pr)\kappa_4 \mathcal{O}_4= -f_1(t_\pr)J_{\rm pr} \, ,
\end{equation*}
where $\mathcal{O}_2$ is the usual Klein-Gordon term 
\begin{equation}
\mathcal{O}_2 = \Box^\pr \pi_\pr \, ,
\end{equation}
and we have two non-linear operators 
\begin{equation}
\mathcal{O}_3 = \left((\Box^\pr \pi_\pr)^2 - (\partial^\pr_\mu \partial^\pr_\nu \pi_\pr)^2 \right) \, ,
\end{equation}
and
\begin{equation}
\mathcal{O}_4 = \left((\Box^\pr\pi_\pr)^3 - 3\Box^\pr\pi_\pr (\partial^\pr_\mu \partial^\pr_\nu \pi_\pr)^2 + 2 (\partial^\pr_\mu \partial^\pr_\nu \pi_\pr)^3 \right). 
\end{equation}
The source is taken to be
\begin{equation}
    J = \frac{2\sqrt{2}}{3\pi} \frac{\Omega \bar{r}}{\sigma_\pr^3} \left(e^{-\left({\vec{r}_+}^{\,\pr}(t)/\sigma_\pr\right)^2}+e^{-\left({\vec{r}_-}^{\,\pr}(t)/\sigma_\pr\right)^2}\right)\, ,
\end{equation}
where we take $\sigma_\pr = \sigma \,2 /\bar{r} = 1/3$.
The cubic interaction is parameterized by the dimensionless quantity
\begin{equation}
\kappa_3 = \frac{1}{3\Lambda_3^3} \sqrt{\frac{2^4M_{\rm s}}{\bar{r}^5}} \, ,
\end{equation}
whereas the quartic interaction is parameterized by
\begin{equation}
\kappa_4 =  \frac{1}{9\Lambda_4^6} \frac{2^4 M_{\rm s}}{\bar{r}^5}.
\end{equation}
If we only consider the cubic term, there is a clear connection between the size of $\kappa_3$ and the Vainstein radius, 
\begin{equation}
  r_{v} = \frac{1}{\Lambda_3}\left(\frac{M_s}{16 \mpl}\right)^{1/3}.
  \label{vradius}
\end{equation}
In the presence of a quartic term, we have two important radii that differentiate between the regions where the quartic term, the cubic term, or the Klein-Gordon term dominates the left hand side of \eqref{eqn:ssymetry}.  While we can set an expectation of where the second of these boundaries is based on \eqref{vradius}, we can numerically solve \eqref{eqn:ssymetry} to find both. \\ 

\begin{figure}
\centering
\includegraphics[width=8cm]{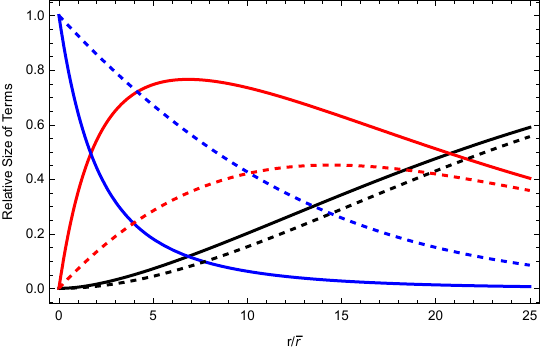}
\caption{\label{fig:comparesizes} The relative contributions of the different terms on the left hand side of the spherically symmetric equation of motion \eqref{eqn:ssymetry}.  We show the relative contributions of the quartic (blue), cubic (red) and Klein-Gordon (black) contributions for our choice of $\xi = 0.6$ (solid) and what would happen in the case of a larger, $\xi = 0.95$ (dashed).  In both cases we set the size of $\kappa_3$ such that the Vainstein radius is approximately $r/\bar{r} = 20$.  }
\end{figure}

In a particular EFT realization, there might be a clear reason to expect that the two interactions scales are related $\Lambda_4 \sim \Lambda_3$, however generically they can be independent. In order to truly probe the Vainshtein screening effect of the quartic Galileon we will chose a value of $\kappa_4$ which is large enough for the quartic term to dominate in the region of the source.  To use the parameterization of \cite{Gerhardinger:2024}, we set
\begin{equation}
\kappa_4 = \kappa_3^2\xi^6 \, ,
\end{equation}
where $\xi = 0.6$ is a fiducial value that realizes our goal.  Fig.~\ref{fig:comparesizes} demonstrates how this choice puts the boundary between the region of quartic dominance and the region of cubic dominate at a value of $r/\bar{r} \approx 1.67$ -- this value is substantially far away from the sources which orbit at $r/\bar{r} = 0.5$ with a width of $\sigma/\bar{r} = 1/6$.
 Larger values of $\xi$ should, in principle, be possible however require significantly more computational time due to the need to slowly turn on the interactions. \\
 
\begin{figure}
\centering
\includegraphics[width=8cm]{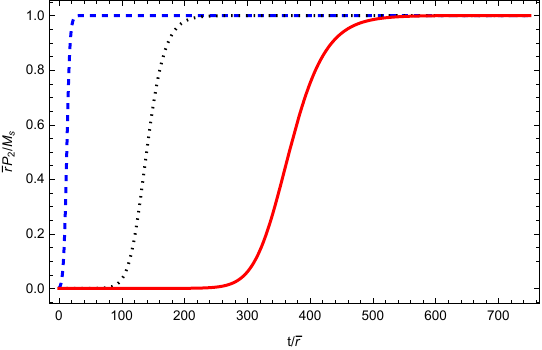}
\caption{\label{fig:turnon} The turn-on functions, $f_1(t)$ (blue, dashed), $f_3(t)$ (black, dotted), and $f_4(t)$ (red solid).  In the fiducial case, $\Omega_p \bar{r} = 0.2$ so the system orbits about every $t/\bar{r} = 30$.}
\end{figure}

 In the present work we need to be more careful than in previous work \cite{Dar:2018dra,Gerhardinger:2022bcw} when turning on the sources and interactions.   These choices are not unique, but they allow us to get all the simulations presented here to a stable state, where the code can run until its endtime.  We turn on the source using 
\begin{equation}
f_1(t_\pr) = \frac{\tanh(0.1(t_\pr- 25)) + \tanh(2.5)}{1 + \tanh(2.5) - 0.01} \, ,
\end{equation}
and the two interactions using 
\begin{equation}
f_3 (t_\pr)=\frac{1}{4}\left(\tanh(.015(t_\pr - 250)) + \tanh(3.75)\right)^2 \, ,
\end{equation}
and
\begin{equation}
f_4(t_\pr ) = \frac{1}{4}\left(\tanh(.0075(t_\pr- 675)) + \tanh(5.0625)\right)^2.
\end{equation}
These are smoother and slower functions than are needed for pure cubic simulations and are shown graphically in  \ref{fig:turnon}.  In addition to the simulations needing longer program time to initialize, we needed to reduce the timestep from the choice made in \cite{Gerhardinger:2022bcw} to $\Delta t = 3\Delta x / 625 = 0.0048 \,\Delta x$.  \\

We use the same CUDA-accelerated version of GABE as described in section III a. of \cite{Gerhardinger:2022bcw}.  In this  numerical scheme, we employ spectral methods to calculate spatial derivatives of $\pi$ and $\dot{\phi}$; because the derivatives exist in Fourier space, we apply a low-pass filter,
\begin{equation}
F(k) = \frac{1}{2}\tanh^{-1}\left(\frac{1}{10}\left[\frac{N}{2}-\frac{k^2}{dk^2}\right]\right)+\frac{1}{2} \, ,
\end{equation}
on $\dot{\phi}$ at the end of every timestep.  As in \cite{Gerhardinger:2022bcw}, the cutoff scale is chosen to be the equivalent 1-dimensional {\sl Nyquist} frequency, $k_{\rm 1DN} = dk\, N/2 = \pi N/2$, where $dk = 2\pi/L$.  Again, this is not a unique choice, but one that allows for our simulations to remain stable until late times. \\

One major challenge that has been apparent in numerical simulations of the Galileon theories is the treatment of out-going wave boundary conditions \cite{Gerhardinger:2022bcw}. In the present work we continue to use 
\begin{equation}
\label{eqn:boundary}
\dot{\pi} = -\frac{\pi}{r} - \partial_r \pi \, ,
\end{equation}
which is strictly speaking only valid for massless Klein-Gordon fields. Although in principle this should be suitable when applied to an interacting Galileon at large distances, in practice it is hard to perform simulations for large hierarchies, meaning that in practice the radii at which we evaluate the boundary conditions is not significantly larger than the Vainshtein radii. Given this, the interaction terms have not truly switched off and the above boundary conditions can lead to waves reflecting at the boundary of the simulation and potentially generating instabilities that terminate the runs too soon. In the alternative UV completion method considered for the cubic Galileon in \cite{Gerhardinger:2022bcw} and the quartic Galileon \cite{Gerhardinger:2024} a second problem arises which is that it is difficult to implement out-going boundary conditions even for a linear massive scalar. We leave to future work a potential solution to these problems.

\section{Results}\label{sec:results}
To illustrate the efficacy of our numerical scheme, we consider the fiducial model with parameters $\Omega_p \bar{r} = 0.2$, $\bar{r}/r_v = 0.05$ together with a choice of $\xi = 0.6$. For these values, our dimensionless couplings are $\kappa_3 = 1.70 \times 10^5$ and $\kappa_4 = 1.35\times 10^9$.  We find that when running the simulations, a large $\xi$ (and hence large quartic Galileon) necessitates a longer turn on time for the simulations, making the runs inefficient. The alternative approach considered in \cite{Gerhardinger:2024} has the advantage that it may be implemented without needing to slowly turn on the interactions. \\

\begin{figure*}[ht!]
\centering
\includegraphics[width=7cm]{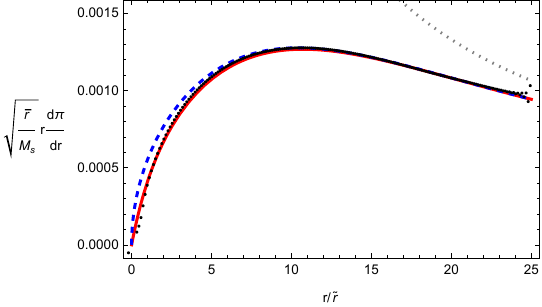}
\includegraphics[width=7cm]{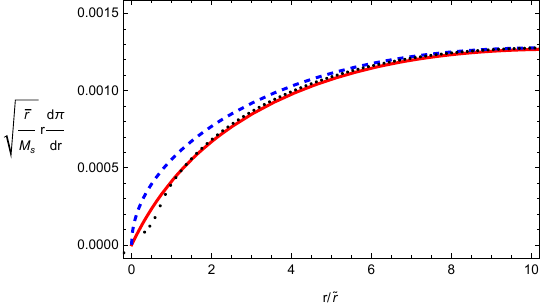}
\caption{\label{fig:rdpidr} The profile of $rE$ averaged over two periods at the end of a simulation with $\Omega_p \bar{r} = 0.2$ (black dots).  This is compared to the static spherically symmetric solution, \eqref{eqn:ssymetry} (red, solid) and a cubic-only static spherically symmetric solution (blue, dotted) and a purely Klein-Gordon static spherically symmetric solution (grey, dashed).  The left panel shows the full range of $0<r<50\bar{r}$ resolved in our simulations and the right panel shows just the range $0<r<20\bar{r}$ to more clearly show disagreement between the spherically symmetric expectations and the simulation near the source.}
\end{figure*}

While there is no analytic solution to the time-dependent problem, we can still look to validate the code, by comparing the time-averaged field profile to that of the (semi-)analytic solution to \eqref{eqn:ssymetry}.  Fig.~\ref{fig:rdpidr} shows a comparison of $r E(r)$, as calculated by taking the positive $x$-axis in the simulation to be the $r$-direction and averaging each point along that axis over the final two periods of the simulation.  We show both the entire range of the positive $x$-axis as well as a close-up near the center.  We can see that the solution is nearly identical to the spherically symmetric profile--and much different from the cubic-only profile.  Deviations from the solution are expected at small $r/\bar{r} \lesssim 1$ where the source is not well described as a point source.

We can now turn to calculating the power emitted by the system.  The outgoing energy flux is given by \cite{Dar:2018dra}
\begin{equation}
t_{0r}^{\pi}=\frac32\left(1+\frac{4}{3\Lambda^{3}}\frac{E}{r}\right)\dot{\pi}\frac{d\pi}{dr}.
\label{eq:outpower}
\end{equation}
Note that this is not the non-linear expression for the power, but rather that for perturbations around a spherically symmetric background. At large distances where the background monopole dominates this is expected to be a sufficiently good approximation to the true radiated power.

\begin{figure}
\centering
\includegraphics[width=8cm]{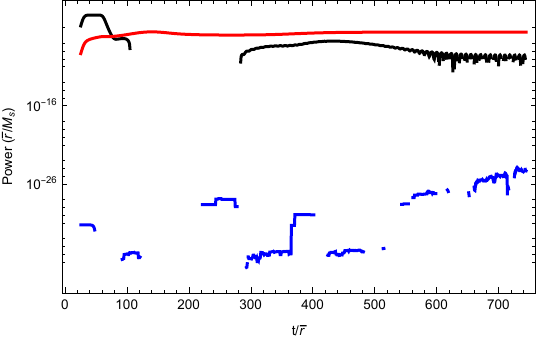}
\caption{\label{fig:polesvstime} Power as a function of time for the fiducial, $\Omega_p\bar{r} = 0.2$ and $\xi = 0.6$ model. 
 The curves show the period-averaged power in the monopole (black), dipole (blue) and quadrupole (red) as a function of time.}
\end{figure}

We evaluate the relation \eqref{eq:outpower} at a radius where the Klein-Gordon term dominates, $r/\bar{r} = 22.5$, which is halfway between the Vainshtein radius and the closest edge of the box.  Again, following the procedure first described in \cite{Gerhardinger:2022bcw}, we evaluate $d \pi/dr$ and $\dot{\pi}$ on a set of points on the sphere defined by the {\sc HEALPIX}\footnote{http://healpix.sourceforge.net} standard using a tri-linear interpolation.  We can then use the efficient {\sc Healpy} \cite{Zonca2019} routines to decompose this power onto the spherical harmonics.  When we report the power, we further perform a rolling time-average over one orbital period.   \\

Fig.~\ref{fig:polesvstime} shows how the first three moments, namely the monopole, dipole and quadrupole, behave as a function of time.  We note that the dipole power is zero to machine precision throughout the simulation. The initially large monopole is an artifact of the way the interactions and source are turned on for which energy is not conserved and a large monopole artificially appears. We find that time-averaged quadrupole power remains the dominant mode, as in Fig.~\ref{fig:polesvstime} in the presence of the quartic interaction. \\

\begin{figure}[H]
\centering
\includegraphics[width=8cm]{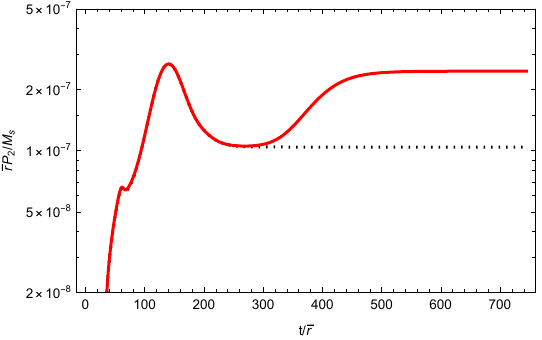}
\caption{\label{fig:quadvscub} The quadrupole power, $P_2$, for our fiducial simulation, $\Omega_p\bar{r} = 0.2$, $\kappa_4 = 0.6$ (red, solid) compared to the quadrupole power in a cubic Galileon, $\kappa_4=0$ simulation (black, dotted).  Note the increase in power as the quartic interactions are turned on.}
\end{figure}

\begin{figure}
\hfill
\subfigure[Power versus multipole. Blue dots denote the cubic, $\kappa_4=0$, Galileon whereas the black dots the fiducial quartic Galileon system.  In both cases, $\Omega_p \bar{r} = 0.2$.  The dashed red line is the best-fit to the low-$\ell$ multipoles for the cubic system and the dashed red line is the best-fit for the quartic system. \label{fig:lowerpoles}]{\includegraphics[width=7.5cm]{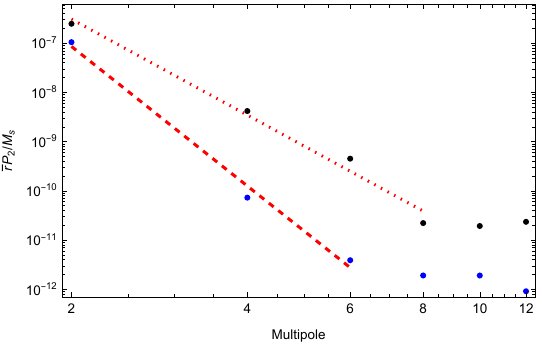}}
\hfill
\subfigure[Power versus orbital velocity. 
 The dots show the the late-time qudrapole power versus $\Omega_p \bar{r}$ for a set of quartic Gaileon simulations. 
 The dotted red line is a best-fit line.\label{fig:powervsalpha}]{\includegraphics[width=7.3cm]{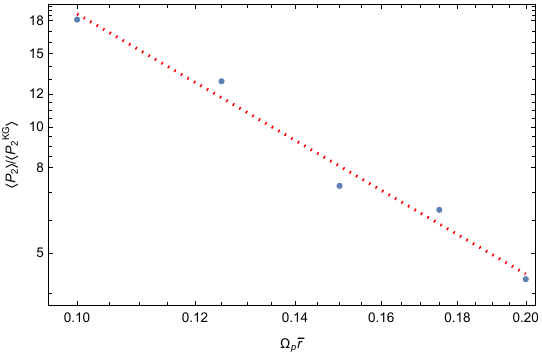}}
\hfill
\caption{The period-averaged power at the end of a set of simulations with (a) multipoles and (b) varying $\Omega_p\bar{r}$.}
\end{figure}

Fig.~\ref{fig:quadvscub} shows how the power in the quadrupole differs in the quartic case from the purely cubic Galileon. Interestingly for our fiducial values the power with the quartic interaction included for the same value of cubic terms is actually larger, although not by orders of magnitude.
Another major difference that arises when the quartic term is added is the modification of the spectrum of power as distributed among the modes. This is illustrated in  Fig.~\ref{fig:lowerpoles}. For cubic case the best-fit shows an $\ell$-dependence of $P\propto \ell^{-6.4756}$ whereas the quartic case shows a best-fit of $P\propto \ell^{-9.414645}$ as in Fig.~\ref{fig:lowerpoles}. \\

Finally, we can now turn our attention to parameterizing how the power depends on $\Omega_p\bar{r}$.  To isolate this, we calculate the ratio of the quadrupole power from our simulations to the Klein-Gordon expectation, \eqref{eq:kgpower}, as can be seen in Fig.~\ref{fig:powervsalpha}.
A simple power-law fit of the scaling of the power with frequency gives
\begin{equation}
    \frac{\left<P_2\right>}{\left<P_2^{\rm KG}\right>} \propto \left(\Omega_p \bar{r}\right)^{-2.07} \, ,
\end{equation}
which varies significantly from the cubic-only case, where \cite{Dar:2018dra}
\begin{equation}
    \frac{\left<P_2^{\rm Cubic}\right>}{\left<P_2^{\rm KG}\right>} \propto \left(\Omega_p \bar{r}\right)^{-2.5} \, ,
\end{equation}
where the latter was consistent with analytic approximations \cite{deRham:2012fw,deRham:2012fg}.
Unfortunately we do not at present have a semi-analytic understanding of this distinct scaling, however it is worth noting that already for the static solutions the scaling of $\pi$ with $r$ in the region where the quartic dominates is quite different from the pure cubic theory.

\section{Conclusions}\label{sec:conclusions}
In this work we have been able to successfully simulate a generic quartic Galileon, and use this to determine the power radiated into scalar radiation for a rotating binary source. This is of great interest in models of modified gravity where additional scalar degrees of freedom can couple to the trace of the stress energy with gravitational strength, but are at the same time screened by the Vainshtein mechanism. Understanding the amount of power for a generic binary system such as a binary pulsar is important in order to put observational constraints on such models. The present work is a step in this direction for modified gravity models whose decoupling limit is well described by a quartic Galileon, as is common in both soft and hard massive gravity theories. \\

First and foremost our results show that the Vainshtein mechanism is fully active in this time-dependent situation, as previously found in simulations of the cubic Galileon \cite{Dar:2018dra,Gerhardinger:2022bcw}. In particular the time averaged field configuration matches well analytic expectations of the screened solution.
This is a non-trivial result since attempts to provide an approximate semi-analytic treatment fail \cite{deRham:2012fw,deRham:2012fg}. We confirm that despite the highly non-linear nature of the system, the dominant scalar radiation is quadrupole, with the next most significant mode $\ell=4$ being typically several orders of magnitude smaller. Although these results parallel the cubic case, we find that the quartic Galileon leads to a qualitatively similar but quantitatively different scaling of the power with orbital velocity and multipole number. \\

Although the particular numerical scheme we have used here is successful, it was necessary to turn each interaction and source on slowly to tame any potential numerical instabilities which may arise due to the fact that the Galileon system is not strictly well-posed. Furthermore the larger the quartic coupling parameter $\xi$ the slower the rate of turn on needed to avoid any instabilities (we were able to simulate $\xi \le 0.6$ with smaller $\xi$ being considerably easier). This unfortunately renders simulating large hierarchies $\xi \gg 1$ too costly in time at present. A solution to the turn on problem is to use the UV completion method proposed in \cite{Gerhardinger:2022bcw} which is considered for the quartic case in \cite{Gerhardinger:2024}. This latter method avoids the particular instabilities associated with the system of equations not being well-posed. Both the present simulation and that of \cite{Gerhardinger:2024} have difficulty giving a proper treatment of the radial boundary conditions and we suspect that a better treatment of the boundary conditions will improve the stability of typical runs.

\section*{Acknowledgements}
We thank Mary Gerhardinger and Mark Trodden for useful discussions and correspondence.
J.T.G.~is supported by the National Science Foundation, PHY-2309919.  The work of CdR and AJT is supported by STFC Consolidated Grant ST/T000791/1 and ST/X000575/1. CdR is also supported by a Simons Investigator award 690508. We acknowledge the National Science
Foundation, Kenyon College and the Kenyon College Department of Physics for providing the hardware used to carry out these simulations.

\bibliographystyle{JHEP}
\bibliography{references}

\end{document}